\documentclass[sigconf, nonacm]{acmart}

\usepackage{typify}
\usepackage{enumitem}
\usepackage{algorithm}
\usepackage{algorithmic}
\usepackage{listings}
\usepackage{makecell}
\usepackage{multirow}

\lstdefinestyle{pythonstyle}{
    language=Python,
    basicstyle=\ttfamily\small,
    keywordstyle=\color{blue},
    stringstyle=\color{magenta!80},
    commentstyle=\color{gray},
    showstringspaces=false,
    columns=flexible,
    numbers=left,
    numberstyle=\small\color{gray},
    stepnumber=1,
    numbersep=6pt,
    frame=single,
    framesep=4pt,
    rulecolor=\color{black!20},
    captionpos=b,
    xleftmargin=5pt,
    xrightmargin=5pt,
    literate={~}{{\textasciitilde}}1,
}
\title{Typify: A Lightweight Usage-driven Static Analyzer for Precise Python Type Inference}

\author{Ali Aman}
\affiliation{%
  \institution{University of Windsor}
  \department{Computer Science}
  \city{Windsor}
  \state{Ontario}
  \country{Canada}
}
\email{burkia@uwindsor.ca}

\author{Muhammad Asaduzzaman}
\affiliation{%
  \institution{University of Windsor}
  \department{Computer Science}
  \city{Windsor}
  \state{Ontario}
  \country{Canada}
}
\email{masaduzz@uwindsor.ca}

\author{Shaowei Wang}
\affiliation{%
  \institution{University of Manitoba}
  \department{Computer Science}
  \city{Winnipeg}
  \state{Manitoba}
  \country{Canada}
}
\email{shaowei.wang@umanitoba.ca}

\begin{document}

\begin{abstract}
Python’s dynamic type system, while offering significant flexibility and expressiveness, poses substantial challenges for static analysis and automated tooling, particularly in unannotated or partially annotated codebases. Existing type inference approaches often depend on existing type annotations, or on deep learning models that require extensive training corpora and considerable computational resources, resulting in limited scalability and reduced interpretability. We introduce \textbf{Typify}, a lightweight, usage-driven static analysis engine designed to infer precise and contextually relevant type information without relying on statistical learning or large datasets. Typify integrates symbolic execution with iterative fixpoint analysis and a context-matching retrieval system to propagate and predict type information across entire projects. By constructing and traversing dependency graphs in an execution-aware manner, Typify accurately connects function calls to their definitions and infers usage-based type semantics, even in complex, interdependent modules. We evaluate Typify on a diverse corpus of real-world Python repositories, including the \textbf{ManyTypes4Py} and \textbf{Typilus} datasets, benchmarking its effectiveness in predicting types of variables, arguments and return statements. Results from the evaluation show that Typify consistently matches or surpasses state-of-the-art deep learning–based systems such as Type4Py and HiTyper, as well as industry standard static type inference tools like Pyre. Our findings demonstrate that usage-driven, retrieval-based inference can match or exceed the accuracy of data-driven methods, offering a practical, interpretable, and computationally efficient alternative for large and evolving Python codebases.

\end{abstract}

\maketitle

\section{Introduction}
Python’s dynamic type system allows developers to write concise and natural code without declaring types for variables, function parameters, or return values. In many cases, especially during rapid prototyping or scripting, omitting type annotations feels more intuitive and keeps code clutter-free. However, this flexibility comes at a cost. Without explicit types, modern development tools such as linters, type checkers, and static analyzers struggle to offer meaningful support~\cite{pep484,hityper}. Bugs that could have been caught early are only revealed at runtime, and navigating or maintaining large untyped codebases becomes significantly more difficult. While Python has officially adopted gradual typing through PEP484~\cite{pep484} and related standards, most real-world code remains largely unannotated. Even in projects where annotations are present, empirical studies show that less than 10\% of annotatable code elements are typed \cite{evo-type-annot}. Manually adding type annotations is tedious and error-prone, especially in existing projects. This has created a growing need for automated type inference tools that can recover missing type information across entire codebases, power static analysis and editor tooling, and generate high-quality annotations without requiring developers to sacrifice flexibility.

Several automated approaches have been proposed to infer or recover types in Python, but each category faces key limitations. \textit{Static analyzers} such as Pytype~\cite{pytype}, Mypy~\cite{mypy}, and Pyright~\cite{pyright} provide full-project analysis but rely heavily on existing annotations and type stubs. They infer only within the narrow bounds of available type annotations and syntax hints, and their reasoning is largely syntax- and constraint-driven rather than usage-driven. In practice, they seldom analyze how functions are called or what values are passed into them, which limits their ability to infer types from call-site behavior. Consequently, when annotations are missing, these tools often fall back to \texttt{Any} or partial inferences, yielding limited coverage for untyped or legacy projects. \textit{Deep learning-based tools} such as Typilus~\cite{typilus} and Type4Py~\cite{type4py} take a data-driven approach, predicting types using neural models trained on large code corpora. While they achieve broad coverage, they are inherently non-deterministic, depend heavily on training data quality, and perform poorly on project-specific or user-defined types. Their predictions are often opaque and difficult to interpret or reproduce, and they require significant computational resources both for training and inference. \textit{Hybrid approaches} like HiTyper~\cite{hityper} attempt to combine static reasoning with learned signals, but they still rely primarily on neural predictions. As a result, they inherit the same generalization and reproducibility issues, while also being limited in scope—typically operating at the file and function level without reasoning across modules or project boundaries.

In this paper, we present \textsc{Typify}, a lightweight and usage-driven static type inference engine for Python that directly addresses these limitations. Unlike prior static tools, \textsc{Typify} does not depend on preexisting annotations or type stubs and is specifically designed to infer types in untyped codebases. It performs deep, usage-aware inference by symbolically simulating code across all reachable paths in a project, tracing function calls, parameter flows, return types, and data structure interactions to infer types directly from observed behavior. This enables \textsc{Typify} to infer types that traditional static analyzers miss—particularly for dynamically created or lazily initialized attributes, which are a common pain point in Python and generally invisible to purely syntactic tools. In contrast to deep learning-based and hybrid approaches, \textsc{Typify} is fully deterministic, requires no training data, and produces interpretable, reproducible results. Furthermore, it performs whole-project reasoning: \textsc{Typify} builds a project dependency graph to analyze modules in dependency order, allowing it to propagate type information across functions, classes, and modules, including user-defined and generic types. When explicit evidence is limited, it complements its usage-driven reasoning with a lightweight retrieval-based context matcher to improve coverage without sacrificing interpretability. By combining cross-module dependency analysis with usage-driven inference and retrieval-based context matcher, we offer a practical solution for high-coverage type inference in real-world Python projects.

We evaluate Typify using two different datasets~\cite{mt4py,typilus}. Results from the study shows that Typify outperforms existing static type checkers. While the technique cannot outperform SOTA hybrid type prediction technique (i.e., HiTyper~\cite{hityper}), Typify stays closely behind of the technique. However, when Typify is combined with a deep learning (DL) model, it outperforms SOTA deep learning models. Compared to other techniques, Typify takes significantly less time in predicting types, making the technique suitable to predict types on the fly or combining with DL models without incurring significant delay.

\noindent\textbf{Contributions.} Our contributions can be summarized as follows:
\begin{itemize}
  \item We propose a purely static type inference framework that integrates \textit{usage-driven} inference for more accurate and context-aware type prediction, without relying on existing type annotations or training data.
  
  \item We develop a robust type inference system, that can incrementally improve type accuracy as it encounters more usage of functions and variables across the project.
  
  \item We implement project-wide inference by building a dependency graph of the entire repository, enabling realistic resolution order and cross-module type recovery, leading to solid improvement compared to existing tools, especially for user-defined types.
  
  \item We perform extensive evaluation of Typify with static type checkers and DL models using ManyTypes4Py~\cite{mt4py} and Typilus~\cite {typilus} dataset.
\end{itemize}

Typify demonstrates that accurate and robust type inference in Python does not require machine learning models or preexisting annotations. Its fully static foundation provides a practical and interpretable path toward building reliable developer tools for dynamic languages. Through symbolic execution, fixpoint reasoning, and whole-project dependency analysis, \textsc{Typify} achieves high coverage and consistency in recovering precise type information across unannotated codebases. These results show that deterministic, explainable, and scalable type inference can be achieved purely through static techniques—bridging the gap between traditional analyzers and learning-based models.

The remainder of this paper is organized as follows. Section~\ref{sec:motivating} presents a motivating example that highlights the limitations of existing approaches and illustrates how \textsc{Typify} overcomes them. Section~\ref{sec:typify} details how types are defined in Python and how Typify represents them. We discuss the design and implementation of our inference engine in this section. Section~\ref{sec:evaluation} describes our experimental setup and evaluates \textsc{Typify} against state-of-the-art baselines using the ManyTypes4Py~\cite{mt4py} dataset and Typilus's~\cite{typilus} dataset. Section~\ref{sec:discuss} discusses limitations, future directions, and threats to validity. Section~\ref{sec:related} reviews related work in type inference and static analysis. Finally, Section~\ref{sec:conclusion} concludes the paper.
\section{Motivating Example}
\label{sec:motivating}

To illustrate the practical gap between existing approaches and usage-driven inference, consider the real-world function shown in Listing~\ref{lst:snippet1}, taken from the \texttt{allennlp} project. The function aggregates string triggers across multiple lists and maps.

\begin{figure}[h]
\centering
\begin{lstlisting}[caption={Example from AllenNLP (simplified for illustration)}, label={lst:snippet1}]
# allennlp/semparse/contexts/atis_tables.py
def get_trigger_dict(seqs, maps):
    merged_map = defaultdict(list)
    for seq in seqs:
        for trigger in seq:
            merged_map[
                trigger.lower()
            ].append(trigger)

    for m in maps:
        for key, values in m.items():
            merged_map[
                key.lower()
            ].extend(values)

    return merged_map

# elsewhere in the project
TRIGGER_LISTS = [["ARRIVE", "DEPART"]]
TRIGGER_DICTS = [{"depart": ["leave", "fly"]}]

get_trigger_dict(TRIGGER_LISTS, TRIGGER_DICTS)
\end{lstlisting}
\end{figure}

The goal is to infer the types of the parameters \texttt{seqs} and \texttt{maps}, and the return type of \texttt{get\_trigger\_dict}.  
This example reflects a challenging pattern in Python: functions whose types are implied only by how they are used elsewhere within the project. The challenge is to determine the origin of the function from its call site, and then cascade the types to the parameters and then within the function as well. 

\paragraph{Why existing methods fail.}
Traditional static analyzers (e.g., Pytype, Pyre) leave the parameters untyped when no explicit annotations are present.  
Unless a type can be trivially inferred from a direct literal assignment (e.g., \texttt{value = 1} $\rightarrow$ \texttt{int}), these tools are unable to infer it.  
Even when the function is used elsewhere with concrete arguments, such information does not propagate back to the function definition—static analyzers treat each function as an isolated unit, and inference stops at the call boundary.

Deep learning–based tools (e.g., Type4Py, Typilus) can predict generic types such as \texttt{list} or \texttt{dict} from token context, but their predictions are probabilistic and typically omit full structure, for instance missing the inner types or mismatching element kinds.

Hybrid systems like HiTyper~\cite{hityper} attempt to combine static and learned reasoning.  
HiTyper models each function as a \emph{Type Dependency Graph (TDG)} where nodes represent variables and edges express data-flow relations.  
Inference proceeds by propagating constraints along these edges and, when local evidence is missing, consulting a neural recommender.  
In this example, both parameters appear as isolated nodes with no incoming edges in the TDG.  
HiTyper therefore asks its neural model for candidate types, but later rejects those recommendations when they do not align with any existing TDG paths.  
Because the TDG is built only at the function level, the tool never considers the external call
\texttt{get\_trigger\_dict(TRIGGER\_LISTS, TRIGGER\_DICTS)}, which already provides concrete evidence about the argument types.  
The result is that all three type slots remain unresolved.

\paragraph{Usage-driven inference.}
Typify addresses these cases by reasoning across the entire project rather than within isolated functions.  
When analyzing the call site, the arguments \texttt{TRIGGER\_LISTS} and \texttt{TRIGGER\_DICTS} have already been visited and typed in earlier passes.  
Instead of querying an external model, Typify looks within the project itself to identify incoming edges by binding the known argument types from each call site to the callee’s parameters. It then proceeds into the function body and applies the same usage-driven reasoning recursively to any additional function calls encountered within.

Through usage-driven propagation, Typify deterministically infers the precise types shown below, with comparative results summarized in Table~\ref{tab:table-1}:
\[
\begin{aligned}
\texttt{seqs} &:\ \texttt{list[list[str]]}, \\
\texttt{maps} &:\ \texttt{list[dict[str, list[str]]]}, \\
\text{return} &:\ \texttt{dict[str, list[str]]}.
\end{aligned}
\]

\begin{table}[t]
\centering
\caption{Prediction results of different baselines for Listing~\ref{lst:snippet1}.}
\label{tab:table-1}
\setlength{\tabcolsep}{1pt}
\renewcommand{\arraystretch}{1}
\small
\begin{tabular}{l p{2.2cm} p{2.2cm} p{2.2cm}}
\toprule
\textbf{Approach} & \textbf{Arg (seqs)} & \textbf{Arg (maps)} & \textbf{Return Type} \\
\midrule
Ground Truth & \texttt{list[list[str]]} & \texttt{list[dict[str, list[str]]]} & \texttt{dict[str, list[str]]} \\
PyType & \texttt{Any} & \texttt{Any} & \texttt{dict[Any, list]} \\
HiTyper & \texttt{Any} & \texttt{Any} & \texttt{Any} \\
Type4Py & \texttt{list[dict[str, Any]]} & \texttt{list[dict[str, Any]]} & \texttt{dict[str, Any]} \\
Typify & \texttt{list[list[str]]} & \texttt{list[dict[str, list[str]]]} & \texttt{dict[str, list[str]]} \\
\bottomrule
\end{tabular}
\end{table}

\paragraph{Summary.}
This example demonstrates a key insight: type information in Python often emerges from \emph{usage} rather than declaration.  
Static analyzers overlook this context, deep learning models approximate it, and hybrid systems like HiTyper cannot extend beyond their local TDGs.  
Typify overcomes these limits by treating the entire project as an interconnected space of type flows—propagating concrete evidence from calls to definitions and producing deterministic, interpretable inferences without annotations or training data.

\medskip
\noindent

\section{Typify}
\label{sec:typify}
In this section, we present the architecture and analysis workflow of Typify, our purely static type inference engine. We first describe its internal type representation, which forms the basis of all inference operations, and then elaborate on the components and algorithms that enable Typify to propagate types, resolve dependencies, and handle advanced typing constructs.

\subsection{Type Representation}
\label{sec:type-repr}
A key design decision in Typify is to represent all types in a uniform way. Before discussing how inference works, we first outline how types are modeled internally.

\begin{figure}[h]
\centering
\begin{tabular}{r p{0.55\linewidth}}
$\theta \in Types$ & $::= \gamma \mid \alpha[\theta, \ldots, \theta] \mid b \mid u \mid \textbf{None} \mid \textbf{type}$ \\
$\gamma \in Elementary~Types$ & $::= \textbf{int} \mid \textbf{float} \mid \textbf{str} \mid \textbf{bool} \mid \textbf{bytes}$ \\
$\alpha \in Generic~Types$    & $::= \textbf{List} \mid \textbf{Tuple} \mid \textbf{Dict} \mid \textbf{Set} \mid \textbf{Callable} \mid \textbf{Generator} \mid \textbf{Union}$ \\
$u \in User\text{-}Defined~Types$ & all user-defined classes, imported types, and named tuples \\
\end{tabular}
\caption{Types in Python.}
\label{fig:type-defs}
\end{figure}

\textbf{TypeExpr.}
The \texttt{TypeExpr} abstraction is Typify’s canonical internal representation of a type. 
It captures the final, fully resolved type of a type slot in a uniform structure that 
is consistent across all type categories. The categories of types handled by Typify are summarized in Figure~\ref{fig:type-defs}. The \texttt{TypeExpr} is equivalent to $\theta$ in the figure.

Conceptually, every type in Typify is expressed as a pair:
\begin{itemize}
    \item \textbf{base} – the underlying class or type definition being represented;
    \item \textbf{args} – a list of nested type expressions representing its parameters.
\end{itemize}

Each argument in \texttt{args} is itself another \texttt{TypeExpr}, allowing recursive composition of 
types without special cases. For example, the type \texttt{dict[str, list[int]]} is represented as a 
single expression whose base is \texttt{Dict} and whose arguments include a string type and a nested 
expression for \texttt{list[int]}.

Typify uses a single uniform abstraction for all kinds of types. 
This consistency makes it straightforward to compare, merge, and transform types during 
inference, since all operations rely on the same internal interface.

Internally, equality and hashing are defined structurally over the base and argument list, 
allowing efficient detection of equivalent inferred types even when they originate from 
different code paths. Normalization routines further ensure that equivalent composite 
forms are treated identically, enabling robust comparison and merging.

Within Typify’s inference pipeline, \texttt{TypeExpr} serves several key purposes:
\begin{itemize}
    \item \textbf{Final storage} — all resolved types in Typify’s symbol tables are stored as \texttt{TypeExpr} instances;
    \item \textbf{Fixpoint detection} — equality checks on \texttt{TypeExpr}s indicate convergence in iterative inference;
    \item \textbf{Result serialization} — \texttt{TypeExpr} can be rendered into a string form for downstream tools such as linters or IDE integrations.
\end{itemize}

In summary, \texttt{TypeExpr} provides a single, recursive structure for representing every possible Python type. 
This uniform representation underpins Typify’s ability to reason about types consistently, 
regardless of their origin or complexity.

\subsection{Overview of the Typify Pipeline}
\label{sec:typify-overview}
\begin{figure*}[t]
  \centering
  \includegraphics[width=0.9\textwidth,keepaspectratio]{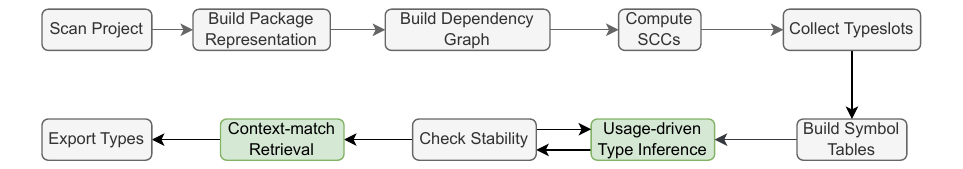}
  \caption{Typify’s inference pipeline.}
  \label{fig:typify-arch}
\end{figure*}

At a high level, Typify processes a Python project through five sequential stages:
\begin{enumerate}
    \item \textbf{Dependency graph construction} (\S\ref{sec:dep-graph})
    \item \textbf{Scheduling and fixpoint resolution} (\S\ref{sec:fixpoint})
    \item \textbf{Usage-driven inference} (\S\ref{sec:module-inference})
    \item \textbf{Context-matching retrieval} (\S\ref{sec:retrieval})
\end{enumerate}
The first three stages constitute Typify’s usage-driven pipeline. The fourth stage is a context-matching retrieval system designed to propose types for slots the usage-driven pipeline could not infer. Figure~\ref{fig:typify-arch} provides a high-level illustration of the process.

\subsection{Dependency Graph Construction}
\label{sec:dep-graph}

Before any inference can begin, Typify constructs a \emph{project-level dependency graph} that captures the import relationships among all modules in the analyzed codebase. This graph forms the structural backbone of the entire inference process: it determines the order in which modules are analyzed, the scope of symbol visibility, and how inferred types are propagated across module boundaries.

Typify begins by scanning the project root and its subpackages to identify all Python modules, including those nested within packages defined by \texttt{\_\_init\_\_.py} files. Each module is assigned a fully qualified name (e.g., \texttt{app.utils.parser}) and represented as a node in the dependency graph. An edge $M_i \rightarrow M_j$ indicates that module $M_i$ directly imports $M_j$. The resulting graph thus encodes the complete import relationships of the project, including third-party and built-in dependencies.

The graph construction process proceeds in three conceptual stages. First, Typify builds an internal package/module level symbol table of the entire project hierarchy. It maps files and directories to Python modules. Second, it traverses each module’s abstract syntax tree (AST), extracting all the import statements to record dependencies. Finally, the discovered edges are inserted into the project dependency graph.

\paragraph{Illustrative example.}
Consider a simple project structure:

\begin{center}
\begin{verbatim}
                  project/
                  |-- __init__.py
                  |-- main.py
                  `-- utils/
                      |-- __init__.py
                      `-- mathops.py
\end{verbatim}
\end{center}

Suppose \texttt{main.py} imports \texttt{utils.mathops}, and \texttt{mathops.py} itself imports Python’s standard \texttt{math} library. When Typify scans this project, it identifies three relevant modules:
\texttt{main}, \texttt{utils.\_\_init\_\_}, and \texttt{utils.mathops}.  
The resulting dependency graph is:

\[
\texttt{main} \rightarrow \texttt{utils.mathops}, \quad
\texttt{utils.mathops} \rightarrow \texttt{math}
\]

Using this dependency graph, Typify derives a topological order for processing the modules:
\[
\texttt{math} \;\rightarrow\; \texttt{utils.mathops} \;\rightarrow\; \texttt{main}
\]
This ensures that any types inferred in \texttt{mathops} are available when analyzing \texttt{main}, maintaining consistency and enabling accurate propagation of inferred information across module boundaries.

\paragraph{Properties.}
The dependency graph is typically sparse and nearly acyclic; however, real-world projects often contain circular imports. Typify therefore detects strongly connected components (SCCs) in the graph and treats each SCC as a unified analysis region. Within an SCC, inference iterates until all module snapshots stabilize, ensuring that types propagated across circular edges converge to a consistent fixpoint. This design enables Typify to perform whole-project reasoning without sacrificing determinism or scalability.

In summary, dependency graph construction establishes the global structure on which Typify’s inference engine operates. By combining AST-based import analysis, absolute-name resolution, and fixpoint iteration over cyclic components, Typify builds a precise and reusable representation of inter-module relationships that guides all downstream type propagation.

\subsection{Scheduling and Fixpoint Resolution}
\label{sec:fixpoint}

Once the dependency graph is built, Typify determines the order in which modules are analyzed. This stage ensures that type information flows consistently across the project and that results remain stable even when modules import each other cyclically.

\paragraph{Scheduling.}
Each module in the dependency graph is scheduled according to its dependencies. For projects without circular imports, the graph is acyclic and modules are processed in \emph{topological order}—a module is analyzed only after all the modules it depends on have been analyzed. This allows inferred types to propagate naturally along import edges, from libraries to utilities to entry points, in a single forward pass.

\paragraph{Fixpoints in cyclic dependencies.}
In real-world projects and especially type stubs, some modules import each other (e.g., \texttt{A} imports \texttt{B} and \texttt{B} imports \texttt{A}). Typify detects such cycles as \emph{strongly connected components} (SCCs) and treats each SCC as a single analysis region. Within a cycle, inference proceeds iteratively until all involved modules stabilize.

Each module maintains a \emph{snapshot} of its current type state—a mapping from all known type slots (variables, parameters, attributes, and returns) to their inferred \texttt{TypeExpr}s. After each iteration, Typify compares the new snapshot to the previous one. If any slot changes, the module is reanalyzed in the next pass; otherwise it is considered stable. The process repeats until no snapshots change, reaching a \emph{fixpoint} where all type information has converged.

\paragraph{Example.}
Consider three modules: $A \rightarrow B \rightarrow C$
where \texttt{A} depends on \texttt{B} and \texttt{B} depends on \texttt{C}. Typify schedules them as:
 $C \;\rightarrow\; B \;\rightarrow\; A$
Since there are no cycles, inference runs once per module, passing inferred types forward through the chain. If instead \texttt{A} and \texttt{B} import each other, Typify groups them into an SCC and alternates analysis: $A \leftrightarrow B$.
The first pass infers partial types, and later passes refine them as new information flows between modules. Once neither module’s snapshot changes, the SCC has reached its fixpoint and inference terminates.

\paragraph{Summary.}
In typical projects, Typify completes inference in a single topological pass. Fixpoint iteration is only required for cyclic dependencies, where it guarantees convergence and consistency across interdependent modules without affecting determinism or performance.

\subsection{Usage-Driven Inference}
\label{sec:module-inference}

Once modules are scheduled, Typify performs \emph{usage-driven inference} within each module.  
This stage is the core of the engine: rather than relying on predefined annotations or training data, Typify infers types directly from how variables, functions, and data structures are \emph{used}.  
The analysis proceeds statement by statement, updating the inferred types of all observable \emph{type slots}—variables, attributes, parameters, and return values—based on evidence collected during traversal. Algorithm~\ref{alg:infermodule} formalizes this process. It presents the statement-by-statement traversal of a module and shows how Typify
propagates inferred types from expressions and call sites into the corresponding type slots.
\paragraph{Overview.}
For each module, Typify first constructs a symbol table recording all locally defined and imported symbols.  
The module body is then traversed in source order.  
For each statement, Typify evaluates its right-hand side (for assignments and expressions) or its call-site context (for function calls) to infer and refine types.  
These inferences are immediately propagated to the corresponding type slots so that later statements can benefit from earlier deductions within the same scope.

\paragraph{Running example.}
Consider listing \ref{lst:usageexample}, representative of many patterns in real-world Python code:

\begin{figure}[h]
\centering
\begin{lstlisting}[caption={Example illustrating usage-driven inference.}, label={lst:usageexample}]
def collect_items(flag):
    items = []                # step 1
    if flag:
        items.append(1)       # step 2
    else:
        items.append("x")     # step 3
    return items
\end{lstlisting}
\end{figure}

When Typify begins analyzing \texttt{collect\_items}, it creates a type slot for each local symbol:  
\texttt{flag}, \texttt{items}, and the function’s return value.

1. Initialization (step 1):  
   The expression \texttt{[]} establishes that \texttt{items} is a list, but its element type is unknown.  
   Typify records the provisional type \texttt{list[Any]}.

2. Usage refinement (steps 2–3):  
   At \texttt{items.append(1)}, Typify observes that the argument \texttt{1} has type \texttt{int}.  
   It refines the slot for \texttt{items} to \texttt{list[int]}.  
   Later, \texttt{items.append("x")} adds a \texttt{str}, and Typify updates the slot again by unifying the observed evidence:  
   \texttt{list[int]} + \texttt{list[str]} → \texttt{list[Union[int, str]]}.  
   This refinement continues throughout the module whenever new, more specific usage is seen.

3. Return propagation:  
   When processing the \texttt{return items} statement, Typify binds the inferred type of \texttt{items} to the function’s return slot, concluding that  
   \[
   \texttt{collect\_items(flag): list[Union[int, str]]}.
   \]
   If the function were later called with arguments whose usage constrained the internal flow further (e.g., only integer appends in observed calls), Typify would refine the type again during project-level propagation.

\paragraph{Refinement and accumulation.}
This incremental refinement mechanism is central to Typify’s precision.  
Whenever a variable, attribute, or parameter participates in a new operation, Typify inspects the operation’s operands to update the slot’s \texttt{TypeExpr}.  
Simple examples include:
\[
\begin{aligned}
x = [] &\Rightarrow x: \texttt{list}, \\
x.append(1) &\Rightarrow x: \texttt{list[int]}, \\
x.append("a") &\Rightarrow x: \texttt{list[Union[int, str]]}, \\
\end{aligned}
\]
Because updates are monotonic—types only become more specific or remain unchanged—analysis remains deterministic and always converges.

\paragraph{Interprocedural reasoning.}
When a call expression is encountered, Typify resolves the callee and infers argument types from the current context.  
These are bound to the callee’s parameters, and the callee’s body is symbolically executed to derive its return type.  
If the same function is invoked with different argument types at distinct sites, Typify merges all observations into the corresponding slots via type unions.  
This ensures that inferred types represent the complete, observed behavior of the function across the project.

\paragraph{Control flow and conservative rules.}
For conditionals and loops, Typify applies sound but conservative reasoning.  
In \texttt{if}/\texttt{else} constructs, type evidence from each branch is merged; in loops, inferred types from the body are unified with their values before the loop.  
This guarantees full coverage of control-flow paths without requiring an explicit control-flow graph.

\paragraph{Summary.}
Usage-driven inference enables Typify to recover rich, context-aware types purely from static evidence.  
Through stepwise refinement, symbolic execution of calls, and monotonic accumulation of observed usage, Typify incrementally constructs an accurate, convergent view of all type slots in a project. These usage-driven inferred types ranked as highest-confidence.  
The next stage (\S\ref{sec:retrieval}) supplements this process with retrieval-based completion for the cases where no usage information is available.

\begin{algorithm}[t]
\caption{\textsc{InferModule}: Usage-driven inference for a given Python file.}
\label{alg:infermodule}
\begin{algorithmic}
\STATE \textbf{Input:} Module $M$
\STATE \textsc{BuildSymbolTable}$(M)$
\FOR{$stmt \in M$ in source order}
  \IF{$stmt$ is \textsc{Assignment}$(x \gets expr)$}
    \STATE $T \gets \textsc{InferExpr}(expr)$
    \STATE \textsc{UpdateTypeSlot}$(M, x, T)$
  \ELSIF{$stmt$ is \textsc{Call}$(target, args)$}
    \STATE $A \gets [\,\textsc{InferExpr}(a) \mid a \in args\,]$
    \STATE $f \gets \textsc{ResolveFunction}(target)$
    \STATE \textsc{BindParameters}$(f, A)$
    \STATE $R \gets \textsc{ExecuteFunction}(f)$
    \STATE \textsc{AccumulateReturnAtCallsite}$(stmt, R)$
  \ELSIF{$stmt$ is \textsc{Return}$(expr)$}
    \STATE $T \gets \textsc{InferExpr}(expr)$
    \STATE \textsc{AccumulateFunctionReturn}$(M, T)$
  \ELSE
    \STATE \textsc{ApplyConservativeRules}$(stmt)$
  \ENDIF
\ENDFOR
\end{algorithmic}
\end{algorithm}

The module pass is straightforward: types flow from observed usage (assignments and calls) to the corresponding type slots, unions summarize variation across call sites, and no speculation is performed when usage is absent. This keeps the analysis predictable and easy to understand while providing high-quality inferences that stabilize under the project-wide fixpoint in \S\ref{sec:fixpoint}, with remaining gaps addressed by the retrieval system in \S\ref{sec:retrieval}.

\subsection{Context-Matching Retrieval}
\label{sec:retrieval}
After usage-driven inference concludes, some type slots may remain unknown (e.g., uncalled functions or sparsely used variables). To improve coverage without sacrificing determinism or interpretability, Typify applies a \emph{context-matching retrieval} step that proposes types for only those empty slots. The key idea is to query a pre-built index of (context, type) pairs extracted from an external knowledge base; retrieved candidates act as suggestions, not overrides. 

\textbf{Retrieval index.}
We build a search index where each entry contains (i) a lightweight context summarizing the slot’s local surroundings and structure, and (ii) the annotated type. For each annotated slot in the training set, we record a small source window around the site; the stored value is the slot’s annotation in string form. Our indexer parses files, for each annotated type slot, it collects the context of surrounding lines, and the annotated type. Using these two items, it writes \texttt{(context, type)} documents into the index.

\textbf{Query construction.}
For each uninferred slot in the analyzed project, Typify constructs a query by obtaining the context of the current type slot. This query is then searched for in the built index. The index returns top-$k$ candidate types whose contexts best match the query. These candidates rank as mid-confidence types. There can still be cases where the query does not return back results. In this scenario we collect all the defined types as well as the imported types in the module, and check if the name of type slot resembles any of the available types, or have common naming conventions associated with common types. This is the lowest-confidence a recommendation can have. Finally, Typify emits the predicted types for each type slot in the project. 

\begin{algorithm}[t]
\caption{\textsc{RetrieveAndFill}: Retrieval-based inference}
\label{alg:retrieval}
\begin{algorithmic}
\STATE {\bf Input:} Module $M$ after usage-driven analysis completes
\FOR{$slot \in \textsc{UninferredTypeSlots}(M)$}
  \STATE $q \gets \textsc{BuildQueryContext}(M, slot)$
  \STATE $C \gets \textsc{SearchIndex}(q)$
  \STATE $T \gets \textsc{SelectType}(C)$
  \IF{$T \neq \bot$}
    \STATE \textsc{WriteTypeIfEmpty}$(slot, T)$
  \ENDIF
\ENDFOR
\end{algorithmic}
\end{algorithm}

\section{Evaluation}

\label{sec:evaluation}
This section discusses the evaluation procedure and the results of our experiments. We answer the following three research questions.
\begin{itemize}
  \item {RQ1: How effective is \emph{Typify} in predicting types compared to deep learning-based type inference systems?}
  \item {RQ2: How effective is \emph{Typify} in predicting types compared to static type inference systems?}
  \item {RQ3: Does the combination of \emph{Typify} with existing type inference systems improve the performance of predicting types?}
\end{itemize}
We run all experiments on Debian 13 LTS with an AMD Ryzen 9 7940HS Processor (@5.2 GHz) and 32 GB RAM.

\subsection{Datasets}
We conducted our experiments using two different datasets. The \emph{ManyTypes4Py} dataset~\cite{mt4py}
consists of Python projects with type annotations. Duplicate files were removed, and the initial dataset was augmented with type annotations using a static type checker, called Pyre \cite{pyre}. For evaluation purposes, we use the test dataset provided as part of the HiTyper replication packages. The original \emph{Typilus} dataset consists of type annotations collected from 600 Python repositories. However, we were able to clone 500 repositories because the remaining repositories are no longer available (e.g., deleted). For the purpose of our evaluation, we consider type annotations from 20\% of the total Python files. However, we did not augment the \emph{Typilus} dataset using any static type checker. A summary of dataset statistics is provided in Table~\ref{tab:dataset-stats}. Please note that type annotations labeled as \texttt{Any} or \texttt{None} are excluded from evaluation, as they do not provide meaningful type information.

\subsection {Evaluation Metrics}
To verify the effectiveness of the compared techniques, we consider two different metrics following prior studies \cite{typilus}. These metrics are defined as follows. 

\textbf{Exact match:} The \textit{exact match} metric denotes the ratio of correct recommendations over all human annotations. Here, a recommendation is considered correct if it matches completely with the human annotation (i.e., the predicted and ground truth types are identical in both base type and type arguments).
\begin{itemize}
        \item \emph{Example:} \texttt{list[str]} vs. \texttt{list[str]} \(\rightarrow\) exact match.
\end{itemize}

\textbf{Base match:} The base match metric is calculated as the ratio of matches over all human annotations without considering the type parameter (i.e, a prediction is considered correct when the predicted and ground truth types are identical in their base types ignoring type parameters).
\begin{itemize}
    \item \emph{Example:} \texttt{list[str]} vs. \texttt{list[int]} \(\rightarrow\) base match.
\end{itemize}

\begin{table}[t]
\centering
\caption{Type distribution in the test sets. “User” indicates user-defined types.}
\label{tab:dataset-stats}
\setlength{\tabcolsep}{1.5pt}
\renewcommand{\arraystretch}{0.9}
\scriptsize
\scalebox{1.1}{
\begin{tabular}{@{}c c c|c c c|c c c@{}}
\toprule
\textbf{Dataset} & \textbf{Category} & \textbf{Total} & \textbf{Variable} & \textbf{Argument} & \textbf{Return} & \textbf{Simple} & \textbf{Generic} & \textbf{User} \\
\midrule
\multirow{2}{*}{\textbf{Type4Py}} 
 & \textbf{Count} & 93{,}850 & 65{,}950 & 18{,}430 & 9{,}470 & 52{,}330 & 20{,}983 & 20{,}537 \\
 & \textbf{Prop.} & 100\% & 70.3\% & 19.6\% & 10.1\% & 55.8\% & 22.4\% & 21.9\% \\
\midrule
\multirow{2}{*}{\textbf{Typilus}} 
 & \textbf{Count} & 93{,}942 & 2{,}076 & 71{,}558 & 20{,}308 & 49{,}888 & 18{,}973 & 25{,}081 \\
 & \textbf{Prop.} & 100\% & 2.2\% & 76.2\% & 21.6\% & 53.1\% & 20.2\% & 26.7\% \\
\bottomrule
\end{tabular}
}
\end{table}

\begin{table*}[!htb]
\centering
\small
\caption{Comparison with the baseline DL approaches considering different tasks.}
\label{tab:rq1-table-1-new}
\renewcommand{\arraystretch}{1.1}
\scalebox{1}{
\begin{tabular}{lll cc|cc|cc}
\toprule
\multirow{3}{*}{\textbf{Dataset}} &
\multirow{3}{*}{\textbf{Task}} &
\multirow{3}{*}{\textbf{Approach}} &
\multicolumn{2}{c|}{\textbf{Top-1}} &
\multicolumn{2}{c|}{\textbf{Top-3}} &
\multicolumn{2}{c}{\textbf{Top-5}} \\
\cline{4-9}
& & &
\multicolumn{1}{c}{\makecell{Exact\\Match}} &
\multicolumn{1}{c|}{\makecell{Base\\Match}} &
\multicolumn{1}{c}{\makecell{Exact\\Match}} &
\multicolumn{1}{c|}{\makecell{Base\\Match}} &
\multicolumn{1}{c}{\makecell{Exact\\Match}} &
\multicolumn{1}{c}{\makecell{Base\\Match}} \\
\hline
\hline

\multirow{12}{*}{\textbf{ManyTypes4Py}} &
  \multirow{3}{*}{\textbf{Variable}} &
    Typify & 62.7 & 74.7 & 71.6 & 82.7 & 73.4 & 84.4 \\
  & & HiTyper & 74.2 & 81.6 & 78.0 & 85.7 & 78.6 & 86.5 \\
  & & Type4Py & 55.2 & 60.1 & 65.9 & 72.6 & 68.1 & 75.1 \\
  \cline{2-9}
  & \multirow{3}{*}{\textbf{Argument}} &
    Typify & 43.0 & 47.7 & 54.1 & 59.0 & 54.5 & 59.4 \\
  & & HiTyper & 45.3 & 47.7 & 55.6 & 58.8 & 56.8 & 60.1 \\
  & & Type4Py & 46.0 & 47.9 & 59.4 & 62.6 & 61.5 & 64.7 \\
  \cline{2-9}
  & \multirow{3}{*}{\textbf{Return}} &
    Typify & 38.4 & 50.5 & 48.2 & 60.3 & 48.6 & 60.7 \\
  & & HiTyper & 46.8 & 56.9 & 51.4 & 61.9 & 52.1 & 62.7 \\
  & & Type4Py & 33.9 & 36.4 & 41.9 & 46.4 & 43.4 & 48.2 \\
  \cline{2-9}
  & \multirow{3}{*}{\textbf{All}} &
    Typify & 55.9 & 66.4 & 65.8 & 75.8 & 67.2 & 77.1 \\
  & & HiTyper & 65.8 & 72.4 & 70.9 & 78.0 & 71.7 & 78.9 \\
  & & Type4Py & 51.2 & 55.3 & 62.2 & 68.0 & 64.3 & 70.4 \\

\hline
\hline

\multirow{12}{*}{\textbf{Typilus's Dataset}} &
  \multirow{3}{*}{\textbf{Variable}} &
    Typify & 18.5 & 55.3 & 18.1 & 61.6 & 18.1 & 61.6 \\
  & & HiTyper & 18.3 & 63.3 & 24.3 & 67.1 & 24.9 & 68.0 \\
  & & Type4Py & 14.4 & 33.6 & 22.8 & 49.0 & 24.9 & 52.4 \\
  \cline{2-9}
  & \multirow{3}{*}{\textbf{Argument}} &
    Typify & 49.2 & 52.2 & 61.2 & 64.6 & 61.4 & 64.8 \\
  & & HiTyper & 52.1 & 54.0 & 62.9 & 65.4 & 64.6 & 67.1 \\
  & & Type4Py & 50.3 & 51.7 & 63.5 & 65.5 & 65.9 & 68.1 \\
  \cline{2-9}
  & \multirow{3}{*}{\textbf{Return}} &
    Typify & 31.6 & 44.0 & 43.5 & 57.1 & 43.8 & 57.5 \\
  & & HiTyper & 40.7 & 52.9 & 44.9 & 57.7 & 45.8 & 58.9 \\
  & & Type4Py & 29.4 & 32.4 & 36.2 & 41.4 & 37.7 & 43.4 \\
  \cline{2-9}
  & \multirow{3}{*}{\textbf{All}} &
    Typify & 44.9 & 50.5 & 56.4 & 62.9 & 56.6 & 63.1 \\
  & & HiTyper & 48.9 & 54.0 & 58.1 & 63.8 & 59.6 & 65.4 \\
  & & Type4Py & 45.0 & 47.1 & 56.7 & 59.9 & 58.9 & 62.4 \\

\bottomrule
\end{tabular}
}
\end{table*}
\subsection{Compared Techniques}
To verify the effectiveness of Typify we consider three baseline approaches (i.e., Type4Py, HiTyper and Pyre Infer). \textbf{Type4Py} is a hierachical neural network model that uses deep similarity learning and type clusters to predict types. \textbf{HiTyper} defines a type dependency graph (TDG) for each function that captures the dependencies among variable. Based on the TDG, the technique uses a combination of static type inference and deep learning approach to predict types. HiTyper uses a set of type rejection rules to eliminate incorrect types. \textbf{Pyre Infer} is a static type inference system for Python developed by Meta.

\label{sec:eval-setup}

\subsection{RQ1: Effectiveness of Typify Compared to Baseline Deep learning-based Approaches}
\label{sec:rq1}
Table~\ref{tab:rq1-table-1-new} presents the comparative results of Typify, HiTyper, and Type4Py on the ManyTypes4Py adataset across variable, argument, and return type prediction tasks.

Overall, HiTyper achieves the best performance across all tasks and metrics, followed closely by Typify, while Type4Py consistently trails behind. For variable type prediction, Typify attains 62.7\% exact-match accuracy and 74.7\% base-match accuracy for top-1 recommendation. Although HiTyper surpasses Typify (74.2\% and 81.6\% respectively), Typify still substantially outperforms Type4Py (55.2\% and 60.1\%). The trend persists for top-3 and top-5 predictions, where Typify achieves 71.6\%/82.7\% (top-3) and 73.4\%/84.4\% (top-5) base-match accuracy. For argument types, Typify shows competitive performance with 43.0\% (exact) and 47.7\% (base) at top-1, and improves to 54.5\%/59.4\% at top-5. HiTyper performs slightly better overall, while Type4Py remains comparable but slightly superior in argument prediction. In return type prediction, Typify’s performance (38.4\% exact, 50.5\% base at top-1) lags behind HiTyper (46.8\% and 56.9\%) but remains ahead of Type4Py. This pattern is consistent for top-3 and top-5 results, where Typify maintains a noticeable advantage over Type4Py. When considering all categories combined, Typify reaches 55.9\%/66.4\% (top-1) and 67.2\%/77.1\% (top-5). Although HiTyper again leads (65.8\%/72.4\% and 71.7\%/78.9\%), Typify demonstrates strong overall performance and consistently exceeds Type4Py (51.2\%/55.3\% and 64.3\%/70.4\%).

Results on the Typilus dataset show a similar trend, with HiTyper maintaining the highest accuracy, followed by Typify and Type4Py. However, the performance gap between Typify and HiTyper narrows, particularly in argument and return type prediction. Although HiTyper performs the best, the result is not surprising because HiTyper uses a combination of static analysis and deep learning approaches. Interestingly, Typify consistently maintains close gap with the best performing model HiTyper without using any deep learning technique.

\begin{figure}[t]
\centering
\includegraphics[width=\columnwidth]{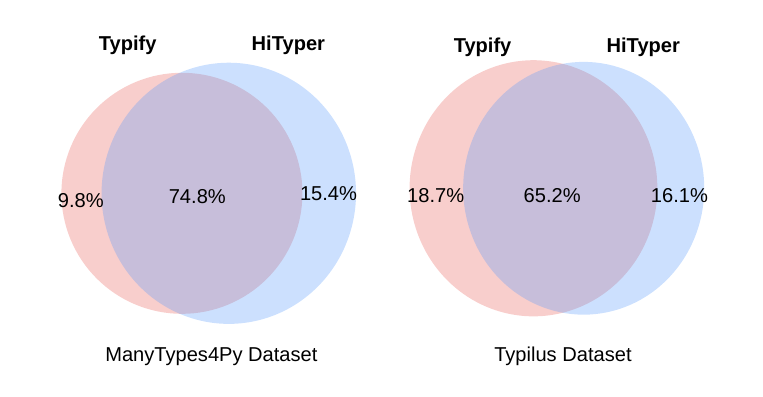}
\caption{Overlap of data points correctly predicted by Typify and HiTyper (Top-1).}
\label{fig:venn-1}
\end{figure}

To determine the extent to which Typify and HiTyper complement each other, we examine the overlap in correctly predicted types, as shown in Figure~\ref{fig:venn-1}. The figure considers only the top-1 predictions for both datasets. Among all data points correctly predicted by either Typify or HiTyper, 9.8\% are predicted exclusively by Typify and 15.4\% exclusively by HiTyper in the \textit{ManyTypes4Py} dataset. For the Typilus dataset, 18.7\% of data points are predicted only by Typify and 16.1\% only by HiTyper. Overall, these results indicate that Typify complements HiTyper. Combining Typify’s recommendations with HiTyper or other deep learning–based approaches can further enhance type inference performance.

\begin{table}[htb]
\centering
\caption{Comparison with the baseline static approaches considering different tasks.}
\label{tab:rq2-table-1}
\renewcommand{\arraystretch}{1.1}
\scalebox{0.85}{
\begin{tabular}{lll cc}
\toprule
\multirow{2}{*}{\textbf{Dataset}} &
\multirow{2}{*}{\textbf{Task}} &
\multirow{2}{*}{\textbf{Approach}} &
\multicolumn{1}{c}{\textbf{Exact}} &
\multicolumn{1}{c}{\textbf{Base}} \\
 & & & \multicolumn{1}{c}{\textbf{Match}} &
       \multicolumn{1}{c}{\textbf{Match}} \\
\hline
\hline

\multirow{12}{*}{\textbf{ManyTypes4Py}} &
  \multirow{3}{*}{\textbf{Variable}} &
    Typify & \textbf{62.7} & \textbf{74.7} \\
  & & HiTyper & 57.0 & 63.8 \\
  & & Pyre Infer & 6.9 & 7.1 \\
  \cline{2-5}
  & \multirow{3}{*}{\textbf{Argument}} &
    Typify & \textbf{43.0} & \textbf{47.7} \\
  & & HiTyper & 7.8 & 8.1 \\
  & & Pyre Infer & 9.4 & 9.4 \\
  \cline{2-5}
  & \multirow{3}{*}{\textbf{Return}} &
    Typify & \textbf{38.4} & \textbf{50.5} \\
  & & HiTyper & 34.0 & 43.7 \\
  & & Pyre Infer & 35.2 & 36.2 \\
  \cline{2-5}
  & \multirow{3}{*}{\textbf{All}} &
    Typify & \textbf{55.9} & \textbf{66.4} \\
  & & HiTyper & 44.2 & 49.9 \\
  & & Pyre Infer & 10.4 & 10.7 \\
\hline
\hline

\multirow{12}{*}{\textbf{Typilus's Dataset}} &
  \multirow{3}{*}{\textbf{Variable}} &
    Typify & \textbf{18.5} & \textbf{55.3} \\
  & & HiTyper & 16.3 & 52.7 \\
  & & Pyre Infer & 0.9 & 1.1 \\
  \cline{2-5}
  & \multirow{3}{*}{\textbf{Argument}} &
    Typify & \textbf{49.2} & \textbf{52.2} \\
  & & HiTyper & 7.0 & 7.1 \\
  & & Pyre Infer & 8.8 & 8.8 \\
  \cline{2-5}
  & \multirow{3}{*}{\textbf{Return}} &
    Typify & \textbf{31.6} & \textbf{44.0} \\
  & & HiTyper & 26.9 & 37.5 \\
  & & Pyre Infer & 27.3 & 28.3 \\
  \cline{2-5}
  & \multirow{3}{*}{\textbf{All}} &
    Typify & \textbf{44.9} & \textbf{50.5} \\
  & & HiTyper & 11.4 & 14.4 \\
  & & Pyre Infer & 12.6 & 12.9 \\

\bottomrule
\end{tabular}
}
\end{table}

\begin{table*}[htb]
\centering
\caption{Performance of the technique that combines Typify with Type4Py (Improvements are shown in the brackets with respect to HiTyper results reported in Table 3).}
\label{tab:rq3-table-1}
\renewcommand{\arraystretch}{1.1}
\scalebox{0.9}{
\begin{tabular}{lll cc|cc|cc}
\toprule
\multirow{3}{*}{\textbf{Dataset}} &
\multirow{3}{*}{\textbf{Task}} &
\multirow{3}{*}{\textbf{Approach}} &
\multicolumn{2}{c|}{\textbf{Top-1}} &
\multicolumn{2}{c|}{\textbf{Top-3}} &
\multicolumn{2}{c}{\textbf{Top-5}} \\
\cline{4-9}
& & &
\multicolumn{1}{c}{\makecell{Exact\\Match}} &
\multicolumn{1}{c|}{\makecell{Base\\Match}} &
\multicolumn{1}{c}{\makecell{Exact\\Match}} &
\multicolumn{1}{c|}{\makecell{Base\\Match}} &
\multicolumn{1}{c}{\makecell{Exact\\Match}} &
\multicolumn{1}{c}{\makecell{Base\\Match}} \\
\hline
\hline

\multirow{3}{*}{\textbf{ManyTypes4Py}} &
  \textbf{Variable} &
    Typify + Type4Py & 75.0 (+0.8) & 86.5 (+4.9) & 78.9 (+0.9) & 90.8 (+5.1) & 79.2 (+0.6) & 91.2 (+4.7) \\
&  \textbf{Argument} &
    Typify + Type4Py & 60.0 (+14.7) & 65.5 (+17.8) & 67.8 (+12.2) & 74.0 (+15.2) & 68.4 (+11.6) & 74.6 (+14.5) \\
&  \textbf{Return} &
    Typify + Type4Py & 48.0 (+1.2) & 60.4 (+3.5) & 54.0 (+2.6) & 67.3 (+5.4) & 54.5 (+2.4) & 67.8 (+5.1) \\

\hline
\hline

\multirow{3}{*}{\textbf{Typilus's Dataset}} &
  \textbf{Variable} &
    Typify + Type4Py & 19.0 (+0.7) & 65.2 (+1.9) & 21.9 (-2.4) & 68.9 (+1.8) & 22.0 (-2.9) & 69.3 (+1.3) \\
&  \textbf{Argument} &
    Typify + Type4Py & 65.4 (+13.3) & 69.4 (+15.4) & 72.9 (+10.0) & 77.1 (+11.7) & 73.1 (+8.5) & 77.5 (+10.4) \\
&  \textbf{Return} &
    Typify + Type4Py & 40.6 (-0.1) & 55.2 (+2.3) & 47.7 (+2.8) & 65.6 (+7.9) & 48.3 (+2.5) & 63.2 (+4.3) \\

\bottomrule
\end{tabular}
}
\end{table*}

\subsection{RQ2: Effectiveness of Typify Compared to Baseline Static Approaches}
\label{sec:rq2}
We analyze Typify’s performance against the static baselines on three prediction tasks (i.e., variables, arguments and return types) across two different datasets. For the purpose of this comparison we only consider the static type checker of the HiTyper. Since, Pyre Infer cannot provide more than one recommendation, thus we compare the result for the top-1 recommendation of Typify and HiTyper. The results are presented in Table~\ref{tab:rq2-table-1}.

For the ManyTypes4Py dataset, Typify consistently outperforms both competing systems, often by large margins. When predicting types of variables, Typify achieves an exact match of 62.7\%, exceeding HiTyper by 5.7 points and Pyre Infer by 55.8 points. When considering the base match, Typify records 74.7\%, which is 10.9 points higher than HiTyper and 67.6 points above Pyre Infer. In case of predicting types of arguments, Typify reaches 43.0\% (exact mach) and 47.7 \% (base match), dramatically surpassing HiTyper (7.8\%, 8.1\%) by 35.2 points and 39.6 points, respectively. Compared with Pyre Infer (9.4\% for both exact and base match), Typify gains +33.6 points in exact match and 38.3 points in base match. When predicting return types,Typify attains 38.4 \% for the exact match and 50.5 \% for the base match, improving upon HiTyper by 4.4 and 6.8 points. Relative to Pyre Infer, Typify is 3.2 points higher in exact match and 14.3 points higher in base match. Overall, Typify outperforming HiTyper by 11.7 and 16.5 points. Against Pyre Infer, the improvements reach 45.5 points and 55.7 points for exact and base matches, respectively.

A similar pattern is observed on the Typilus dataset, reinforcing Typify’s robustness across diverse code distributions. When predicting types of variables, Typify achieves an exact match of 18.5 \% and a base match of 55.3 \%, exceeding HiTyper by 2.2 and 2.6 points, and outperforming Pyre Infer by 17.6 and 54.2 points. For predicting types of arguments, Typify’s results of 49.2 \% exact match and 52.2 \% base match constitute substantial improvements over HiTyper (7.0 \%, 7.1 \%), and over Pyre Infer (8.8\%, 8.8 \%). While predicting return types Typify obtains 31.6 \% of exact match and 44.0 \% of base match, outperforming HiTyper by +4.7 and +6.5, and Pyre Infer by +4.3 and +15.7 points. Overall, Typify performs the best compared to the other two techniques. 

\subsection{Results for RQ3: Combining Typify with Deep learning-based Approaches}
\label{sec:rq3}
For the purpose of this study, we combine the recommendation of Typify with the recommendation obtained from a SOTA deep DL model, called Type4Py. In practice, any DL model can be combined with Typify. We are interested in learning whether combining Typify with a DL model can help us improve the performance of predicting types. As discussed earlier, Typify produces 3 levels of confidence for its predicted types; high, mid, and low. In order to perform this experiment, whenever Typify encounters a type slot for which it has low-confidence level, we use the recommendation from Type4Py. The results are presented in Table \ref{tab:rq3-table-1}. For variable type prediction, Typify+Type4Py achieved a top-1 exact-match accuracy of 75.0\% (86.5\% for the base match), which higher than the result of HiTyper or Type4Py alone. Similar trends were observed in Top-3 and Top-5 recommendation, where Typify+Type4Py outperformed Type4Py or HiTyper. In argument type prediction, Typify+Type4Py achieved 60.0\% Top-1 exact match (65.5\% for the base match). The Top-5 base-match accuracy reached 74.6\%, again exceeding HiTyper and Type4Py. For return types, Typify+Type4Py recorded 48.0\% Top-1 exact match (60.4\% for the base match). Its Top-5 base-match accuracy of 67.8\% reflects higher result than that of HiTyper and Type4Py. We also observe similar results for the Typilus dataset. For example, for the argument type prediction, Typify+Type4Py achieved an exact match of 69.4\% for the top-1 recommendation, and the number increases to 73.1\% for the top-5 recommendation. Thus, we find that DL type inference techniques can easily incorporate Typify to improve the overall performance of type predictions.

\section{Discussion}
\label{sec:discuss}
This section discusses questions related to our study.

\subsection{Timing Efficiency}
We evaluated the runtime efficiency of Typify, HiTyper, and Type4Py to assess their practicality for large-scale or real-time type inference applications. All techniques were benchmarked on the same hardware configuration. On the ManyTypes4Py dataset, Typify required an average of 4.7 ms per data point, compared to 48.2 ms for HiTyper and 36.5 ms for Type4Py. This corresponds to a 90.3\% reduction in latency relative to HiTyper and an 87.1\% reduction relative to Type4Py. Similar trends were observed on the Typilus dataset, where Typify achieved an average inference time of 3.9 ms per data point, compared with 45.3 ms for HiTyper and 33.5 ms for Type4Py, corresponding to 91.4\% and 88.4\% reductions respectively. These results indicate that Typify’s architecture achieves faster prediction without compromising accuracy. Since Typify is a static approach, we ignore the training time required for Type4Py in this comparison.

\subsection{Effectiveness in Predicting Rare Types}
\emph{Rare types} as defined as those representing less than 0.1\% of all annotated types in the dataset. In ManyTypes4Py, 41.3\% of these rare types are generic and 58.4\% are user-defined. Similarly, in Typilus’s dataset, 50.8\% are generic and 48.5\% are user-defined. When integrated with Type4Py, Typify achieves approximately 39\% coverage of rare types, outperforming both HiTyper (33\%) and Type4Py alone (20\%), all under top-1 exact match metric. These findings indicate that combining our usage-driven inference with a recommendation-based model such as Type4Py substantially improves the prediction of rare types.

\subsection{Limitations of the Proposed Technique}
Several limitations follow from Typify’s evidence-first design and retrieval scope:
\begin{itemize}
    \item \textbf{Uncalled functions and incomplete evidence.}
    If a function is never invoked, usage-driven inference cannot constrain its parameters or returns. Typify still infers self-contained returns:
\begin{lstlisting}[xleftmargin=2em]
def default_opts():
    return {"timeout": 10, "verbose": False}
\end{lstlisting}
    but remains agnostic when logic depends on unknown parameters:
\begin{lstlisting}[xleftmargin=2em]
def transform(x):
    return x.to_dict()    # 'x' unknown
\end{lstlisting}
    In these cases, \emph{context-matching retrieval} may fill gaps when the local context resembles indexed examples; otherwise, slots remain unknown.

    \item \textbf{Retrieval mismatch and bias.}
    The retrieval index reflects the files it was built on. Domain shift (e.g., project-specific idioms, novel libraries) can yield weak matches or inaccurate suggestions. Typify mitigates this by never overriding usage-driven types, and always giving preference to the usage-driven type in case of a conflict.

    \item \textbf{Dynamic constructs and runtime indirection.}
    Highly dynamic patterns (reflection via \texttt{getattr}, dynamic imports, code generation, deserialization) remain challenging for purely static reasoning and are not reliably addressed by retrieval unless the surrounding static context is distinctive and well represented in the index.
\end{itemize}

\section{Related Work}

\label{sec:related}
\balance

\textbf{Static vs.\ dynamic inference for Python.}
Existing static type checkers for Python---including \textsc{Mypy}, \textsc{Pyre}, \textsc{Pytype}, \textsc{Pyright}, \textsc{PySonar2}, and \textsc{Jedi}---are primarily designed for type checking rather than type inference~\cite{mypy,pyre,pytype,pyright,pysonar2,jedi}. While they are effective at inferring types in codebases that are already heavily annotated, they struggle with unannotated code, reflective features, and dynamic import patterns. Design choices in the Python typing ecosystem (e.g., PEP~484 for type hints, PEP~526 for variable annotations, PEP~561 for the distribution of stub packages, deferred or stringized annotations in PEP~563, and the recent generic syntax in PEP~695) have substantially shaped both the opportunities and constraints for inference and checking in practice~\cite{pep484,pep526,pep561,pep563,pep695}. Beyond local reasoning, static call-graph construction is a key enabler for interprocedural inference; \textsc{PyCG} demonstrates a scalable approach for Python that handles higher-order functions and import graphs~\cite{pycg}.

\textbf{Hybrid and dynamic approaches.}
Dynamic and hybrid techniques trade coverage for precision by executing programs under representative inputs and propagating observed types, or by integrating runtime signals into static analyses. \textsc{HiTyper} explicitly weaves a rule-based static engine with DL predictions over type-dependency graphs, using rejection rules to guard correctness \cite{hityper}. Related efforts show that purely dynamic or hybrid checkers can improve precision but may face scalability and input-dependence limits on large code bases \cite{mypy,pyre,pytype,pyright}.

\textbf{Learning-based type inference.}
Learning-based methods frame type inference as prediction. Probabilistic approaches rank candidate types to handle ambiguity \cite{xu2016probpy}. Neural sequence models such as \textsc{DeepTyper} leverage local context \cite{hellendoorn2018deeptyper}, while graph-based systems like \textsc{Typilus} exploit structural information to recover rare and user-defined types \cite{typilus}. For JavaScript/TypeScript, \textsc{LambdaNet} and \textsc{TypeBERT} apply GNNs and large-scale pretraining to type prediction~\cite{lambdanet,typebert}. In Python, \textsc{TypeWriter}, \textsc{PYInfer}, and \textsc{Type4Py} target function and variable types using neural inference, search, or similarity learning \cite{typewriter,pyinfer,type4py}. Recent work explores LLM-based generation with static-knowledge prompting (\textsc{TypeGen}) and systematic evaluation of end-to-end annotation quality \cite{typegen,generate-annotations}. Benchmarks such as \textsc{TypeEvalPy} enable reproducible evaluation of precision, recall, and rare-type coverage \cite{typeevalpy}.

\textbf{Type inference beyond Python.}
Statically typed languages (e.g., Java) use mostly local inference for usability and performance \cite{pierce2000local}, supported by interprocedural analyses like CHA, RTA, and points-to analysis \cite{cha,rta,andersen94,steensgaard96}. For dynamic languages, gradual typing systems (e.g., TypeScript, Flow) combine inference, constraint solving, and library stubs \cite{typescript,flow}, while Typed Racket offers a principled model with blame tracking \cite{typedracket}.

\textbf{Positioning.}
Compared to DL-only systems \cite{type4py,hellendoorn2018deeptyper,typilus} and hybrid frameworks like \textsc{HiTyper} \cite{hityper}, our usage-driven static analyzer augments analysis with a retrieval layer to enhance the completion of type inference. This helps generalize to rare and user-defined types without requiring any dynamic execution, while remaining compatible with the evolving Python typing standards.
\section{Conclusion}
\label{sec:conclusion}
We introduced Typify, a purely static, fine-grained type inference engine for Python that achieves context-sensitive precision without runtime execution or deep learning. Typify builds on a uniform internal representation (\texttt{TypeExpr}) and employs a fixpoint-based analysis pipeline to propagate type information across modules, functions, and recursive call structures until convergence. By explicitly modeling program state, dependencies, and generics, Typify achieves high accuracy and robustness, even for projects with complex import hierarchies or incomplete annotations. Our evaluation shows that Typify performs competitively with, and often surpasses, state-of-the-art deep learning systems such as Type4Py, while matching or exceeding hybrid approaches like HiTyper. Moreover, combining Typify with a learned model (e.g., Type4Py) further improves accuracy, outperforming existing hybrid solutions. Beyond predictive performance, Typify’s location-aware type slots enable precise downstream applications like static checking and identifying buggy code locations. Despite its expressiveness, Typify remains efficient: its static architecture scales to multi-module codebases without requiring annotations or training data.

Future work includes enhancing control-flow and alias analyses, improving alias normalization for more consistent equivalence resolution, and advancing retrieval through learning-to-rank strategies, domain-specific sub-indexing, and project-adaptive indexing. Another key direction is formalizing Typify into a practical, developer-facing tool to facilitate real-world integration and usability. Together, these extensions aim to broaden Typify's coverage—particularly for uninvoked functions and sparsely exercised modules—while preserving its evidence-driven inference foundations. The replication package for this study is publicly available on GitHub at \textbf{\url{https://github.com/ali-aman-burki/typify}}.

\begin{acks}
The authors would like to thank the anonymous reviewers for their valuable feedback.
This research is supported in part by the Natural Sciences and Engineering Research Council of Canada (NSERC) Discovery Grants program.
\end{acks}

\bibliographystyle{plain}  
\bibliography{refs}

@inproceedings{evo-type-annot,
  author    = {L. Di Grazia and M. Pradel},
  title     = {The evolution of type annotations in Python: an empirical study},
  year      = {2022},
  booktitle = {Proceedings of the 30th ACM Joint European Software Engineering Conference and Symposium on the Foundations of Software Engineering (ESEC/FSE)},
  pages     = {209--220}
}

@inproceedings{hityper,
  author    = {Y. Peng and C. Gao and Z. Li and B. Gao and D. Lo and Q. Zhang and M. Lyu},
  title     = {Static inference meets deep learning: a hybrid type inference approach for Python},
  year      = {2022},
  booktitle = {Proceedings of the 44th International Conference on Software Engineering (ICSE)},
  pages     = {2019--2030}
}

@inproceedings{typilus,
  author    = {M. Allamanis and E. T. Barr and S. Ducousso and Z. Gao},
  title     = {Typilus: neural type hints},
  year      = {2020},
  booktitle = {Proceedings of the 41st ACM SIGPLAN Conference on Programming Language Design and Implementation (PLDI)},
  pages     = {91--105}
}

@inproceedings{type4py,
  author    = {A. M. Mir and E. Lato{\v{s}}kinas and S. Proksch and G. Gousios},
  title     = {Type4Py: practical deep similarity learning-based type inference for Python},
  year      = {2022},
  booktitle = {Proceedings of the 44th International Conference on Software Engineering (ICSE)},
  pages     = {2241--2252}
}

@inproceedings{mt4py,
  author    = {A. M. Mir and E. Lato{\v{s}}kinas and G. Gousios},
  title     = {ManyTypes4Py: a benchmark Python dataset for machine learning-based type inference},
  year      = {2021},
  booktitle = {Proceedings of the IEEE/ACM 18th International Conference on Mining Software Repositories (MSR)},
  pages     = {585--589}
}

@misc{pep484,
  author       = {G. van Rossum and J. Lehtosalo and {\L}. Langa},
  title        = {PEP 484: type hints},
  year         = {2014},
  howpublished = {\url{https://peps.python.org/pep-0484/}},
  note         = {Retrieved Jan 29, 2026}
}

@misc{pyright,
  title        = {Pyright},
  year         = {2026},
  howpublished = {\url{https://microsoft.github.io/pyright/}},
  note         = {Retrieved Jan 29, 2026}
}

@misc{pyre,
  title        = {Pyre},
  year         = {2026},
  howpublished = {\url{https://pyre-check.org/}},
  note         = {Retrieved Jan 29, 2026}
}

@misc{pytype,
  title        = {Pytype},
  year         = {2026},
  howpublished = {\url{https://google.github.io/pytype/}},
  note         = {Retrieved Jan 29, 2026}
}

@misc{mypy,
  title        = {Mypy},
  year         = {2026},
  howpublished = {\url{https://mypy-lang.org/}},
  note         = {Retrieved Jan 29, 2026}
}

@inproceedings{hellendoorn2018deeptyper,
  author    = {V. J. Hellendoorn and C. Bird and E. T. Barr and M. Allamanis},
  title     = {Deep learning type inference},
  year      = {2018},
  booktitle = {Proceedings of the 26th ACM Joint European Software Engineering Conference and Symposium on the Foundations of Software Engineering (ESEC/FSE)},
  pages     = {152--162}
}

@inproceedings{lambdanet,
  author    = {J. Wei and M. Goyal and R. Jain and B. Nieuwenhuis and H. Madhyastha and P. Anderson and I. Dillig},
  title     = {LambdaNet: probabilistic type inference using graph neural networks},
  year      = {2020},
  booktitle = {International Conference on Learning Representations (ICLR)}
}

@inproceedings{typebert,
  author    = {K. Jesse and V. Raychev and M. Pradel and P. Devanbu},
  title     = {Learning type annotation: is big data enough?},
  year      = {2021},
  booktitle = {Proceedings of the 29th ACM Joint European Software Engineering Conference and Symposium on the Foundations of Software Engineering (ESEC/FSE)}
}

@inproceedings{xu2016probpy,
  author    = {Z. Xu and V. Raychev and M. Vechev and T. Touili},
  title     = {Python probabilistic type inference with natural language support},
  year      = {2016},
  booktitle = {Proceedings of the ACM SIGPLAN International Conference on Generative Programming: Concepts and Experiences (GPCE)}
}

@inproceedings{typewriter,
  author    = {M. Pradel and G. Gousios and J. Liu and S. Chandra},
  title     = {TypeWriter: neural type prediction with search-based validation},
  year      = {2020},
  booktitle = {Proceedings of the 28th ACM Joint European Software Engineering Conference and Symposium on the Foundations of Software Engineering (ESEC/FSE)},
  pages     = {209--220}
}

@inproceedings{pyinfer,
  author    = {S. Cui and L. Zhao and X. Li and J. Huang},
  title     = {PYInfer: deep learning semantic type inference for Python variables},
  year      = {2020},
  booktitle = {Proceedings of the 35th IEEE/ACM International Conference on Automated Software Engineering (ASE) Workshops}
}

@article{typegen,
  author  = {Y. Peng and C. Gao and Z. Li and D. Lo and M. Lyu},
  title   = {Generative type inference for Python},
  year    = {2023},
  journal = {arXiv}
}

@inproceedings{pycg,
  author    = {V. Salis and T. Sotiropoulos and P. Louridas and D. Spinellis and D. Mitropoulos},
  title     = {PyCG: practical call graph generation in Python},
  year      = {2021},
  booktitle = {Proceedings of the 43rd International Conference on Software Engineering: Companion Proceedings (ICSE-Companion)}
}

@misc{pep526,
  author       = {J. Lehtosalo},
  title        = {PEP 526: syntax for variable annotations},
  year         = {2016},
  howpublished = {\url{https://peps.python.org/pep-0526/}},
  note         = {Retrieved Jan 29, 2026}
}

@misc{pep561,
  author       = {J. Lehtosalo and G. van Rossum},
  title        = {PEP 561: distributing and packaging type information},
  year         = {2017},
  howpublished = {\url{https://peps.python.org/pep-0561/}},
  note         = {Retrieved Jan 29, 2026}
}

@misc{pep563,
  author       = {L. Hastings},
  title        = {PEP 563: postponed evaluation of annotations},
  year         = {2017},
  howpublished = {\url{https://peps.python.org/pep-0563/}},
  note         = {Retrieved Jan 29, 2026}
}

@misc{pep695,
  author       = {E. Traut and J. Zijlstra},
  title        = {PEP 695: type parameter syntax},
  year         = {2023},
  howpublished = {\url{https://peps.python.org/pep-0695/}},
  note         = {Retrieved Jan 29, 2026}
}

@misc{pysonar2,
  author       = {Y. Wang},
  title        = {PySonar2},
  year         = {2026},
  howpublished = {\url{https://github.com/yinwang0/pysonar2}},
  note         = {Retrieved Jan 29, 2026}
}

@misc{jedi,
  title        = {Jedi},
  year         = {2026},
  howpublished = {\url{https://jedi.readthedocs.io/}},
  note         = {Retrieved Jan 29, 2026}
}

@article{generate-annotations,
  author  = {Y. Zhang et al.},
  title   = {Generating Python type annotations from type inference},
  year    = {2024},
  journal = {Communications of the ACM}
}

@article{typeevalpy,
  author  = {A. P. S. Venkatesh and R. L{\"a}mmel and E. Bodden},
  title   = {TypeEvalPy: a micro-benchmarking framework for Python type inference and checking},
  year    = {2023},
  journal = {arXiv}
}

@article{pierce2000local,
  author  = {B. C. Pierce and D. N. Turner},
  title   = {Local type inference},
  year    = {2000},
  journal = {ACM Transactions on Programming Languages and Systems},
  volume  = {22},
  number  = {1},
  pages   = {1--44}
}

@inproceedings{cha,
  author    = {J. Dean and D. Grove and C. Chambers},
  title     = {Optimization of object-oriented programs using static class hierarchy analysis},
  year      = {1995},
  booktitle = {Proceedings of the European Conference on Object-Oriented Programming (ECOOP)},
  pages     = {77--101}
}

@inproceedings{rta,
  author    = {D. F. Bacon and P. F. Sweeney},
  title     = {Fast static analysis of C++ virtual function calls},
  year      = {1996},
  booktitle = {Proceedings of the ACM Conference on Object-Oriented Programming, Systems, Languages, and Applications (OOPSLA)},
  pages     = {324--341}
}

@techreport{andersen94,
  author      = {L. O. Andersen},
  title       = {Program analysis and specialization for the C programming language},
  year        = {1994},
  institution = {DIKU, University of Copenhagen},
  number      = {148}
}

@inproceedings{steensgaard96,
  author    = {B. Steensgaard},
  title     = {Points-to analysis in almost linear time},
  year      = {1996},
  booktitle = {Proceedings of the ACM SIGPLAN-SIGACT Symposium on Principles of Programming Languages (POPL)},
  pages     = {32--41}
}

@misc{typescript,
  title        = {TypeScript},
  year         = {2026},
  howpublished = {\url{https://www.typescriptlang.org/}},
  note         = {Retrieved Jan 29, 2026}
}

@misc{flow,
  title        = {Flow},
  year         = {2026},
  howpublished = {\url{https://flow.org/}},
  note         = {Retrieved Jan 29, 2026}
}

@inproceedings{typedracket,
  author    = {S. Tobin-Hochstadt and M. Felleisen},
  title     = {Typed Racket: a practical gradual type system},
  year      = {2008},
  booktitle = {Proceedings of the ACM International Conference on Functional Programming (ICFP)},
  pages     = {78--89}
}

\end{document}